\documentstyle[12pt]{article}
\def\k{{\rm {\bf k}}}
\def\p{{\rm {\bf p}}}
\def\q{{\rm {\bf q}}}
\def\r{{\rm {\bf r}}}

\def\dij{\delta_{ij}}

\begin{document}

\hfill BI-TP 94/08

\hfill February 1994

\vspace{1.5cm}

\begin{center}
{\bf THE NONABELIAN SCREENING POTENTIAL BEYOND THE LEADING ORDER }

\vspace{1.5cm}
{\bf R.~Baier and O.~K.~Kalashnikov}
\footnote{Permanent address: Department of
Theoretical Physics, P.N.Lebedev Physical Institute, Russian
Academy of Sciences, 117924 Moscow, Russia. E-mail address:
kalash@td.fian.free.net}

Fakult\"at f\"ur Physik

Universit\"at Bielefeld

D-33501 Bielefeld, Germany

\vspace{2.5cm}
{\bf Abstract}
\end{center}

The nonabelian screening potential is calculated
 in the temporal axial gauge.
The Slavnov-Taylor identity is used to construct the three-gluon
vertex function from the inverse gluon propagator.
After solving
 the Schwinger - Dyson equation beyond
leading order we find that the obtained momentum dependence of the
gluon self-energy at high temperature does not correspond to
an attractive QCD - Debye  potential,
but instead it is repulsive and power  behaved
($  \simeq 1/r^6$) at large distance.

\newpage

\leftline{\bf  1  Introduction}
\bigskip

In recent papers [1,2] Rebhan has investigated the
screening (Debye)  mass for QCD at finite temperature
 beyond the leading order, especially under the aspect
of its gauge dependence (in the case of covariant gauges).
This analysis has been  performed in the framework of the
 Braaten - Pisarski
 resummation scheme [3], and it is successful to define
 a gauge-independent screening length
in  next-to-leading order.
However, it is logarithmically sensitive to the momentum scale
$O(g^2 T)$, e.g. to the nonperturbative
gluon magnetic mass, and it differs from previously found
results [4,5].

On the other hand
possible modifications of Debye screening have been discussed in
 Refs.[6,7],  where it is shown that the momentum dependence of the
 gluon self-energy may essentially affect the large distance behaviour
of the potential.

In  this paper we calculate the static QCD coulor-singlet
 potential beyond the
leading order by exploiting the  infrared limit of the
(time-time component of the) gluon self-energy
 $\Pi_{44}(0,\k)$.
We work in
the temporal axial gauge, and we use the full Schwinger-Dyson
  equation for $\Pi_{44}$, which includes the
 ring graphs as well as  nonperturbative vertex corrections.
Gauge covariance is guaranteed by  the exact
Slavnov-Taylor identities in order
 to construct the  three-gluon vertex [8,9].

 As a result  we find  strong indications
for a power-like screening potential  for
 high temperature QCD.

\bigskip
\leftline{\bf 2 The gluon polarization tensor and the three-gluon vertex}
\bigskip

The following calculations are performed
in the temporal axial gauge [4], which is well suited to
keep gauge covariance and to evaluate the necessary infrared limit.
In this gauge, in which the gauge vector is parallel to the
velocity vector $u_{\mu}$ of the medium,
 the thermal Green function
technique is
considerably simplified, especially since
 the exact gluon propagator has only non-vanishing spatial components,
$$
 {\cal D}_{ij} (k) =
\frac{1}{k^2+G}\left(\delta_{ij}-\frac{k_ik_j}{\k^2}\right)+
\frac{1}{k^2+F}\frac{k^2}{k_4^2}\frac{k_ik_j}{\k^2} ,
\eqno{(1)}
$$
 where we  suppress the  colour indices.
${\cal D}_{ij}$
 only depends on two
 scalar functions $F(k)$ and $G(k)$, which are related to the
 gluon polarization tensor $\Pi_{\mu \nu}$ as
follows
$$
G(k)=\frac{1}{2}\left(\sum_{i}\Pi_{ii}+
\frac{k_4^2}{\k^2}\Pi_{44}\right)\,,\qquad
F(k)=\frac{k^2}{\k^2}\Pi_{44}\,.\
\eqno{(2)}
$$
A further important advantage of the axial gauge is provided by
the exact Slavnov-Taylor identity for the three-gluon vertex
function in the form
$$
r_\mu\Gamma_{\mu\nu\gamma}^{abc}(r,\,p,\,q)=igf^{abc}
[{\cal D}_{\nu\gamma}^{-1}(p)-{\cal D}_{\nu\gamma}^{-1}(q)]\,,\
\eqno{(3)}
$$
where $f^{abc}$ denote the structure constants of the colour group.

This equation being rather simple is used to construct the
nonperturbative vertex function necessary for calculating
the gluon-self energy in the infrared limit, namely the
$\Pi_{44}(0,|\q|)$ function,
which is required to determine the static screening  potential.

The exact nonperturbative  graph representation for this tensor
is well-known (see e.g. Refs.[4,9]). In the axial gauge
it consists
of  four diagrams
 for the pure gluon contribution.
 However, when one considers
the $\Pi_{44}$-component only  two nonperturbative diagrams
contribute (the more complicated ones are vanishing).
At finite temperature $T = 1/\beta$
the analytical expression for these  two nonvanishing graphs
is found to be
\setcounter{equation}{3}
\begin{eqnarray}
\Pi_{44}(0,|\q|)&=&
\frac{g^2N}{\beta}\sum_{p_4}
\int\frac{d^3\p}{(2\pi)^3}{\cal D}_{ii}(p)\\
&&-\frac{g^2N}{2\beta}\sum_{p_4}
\int\frac{d^3\p}{(2\pi)^3}
(2p_4)\left[{\cal D}_{li}(p+q)
\Gamma_{ij4}(p+q,\, -p,\, -q){\cal D}_{jl}(p)\right]     \, ,
\nonumber\end{eqnarray}
where
 $\Gamma_{ij4}^{abc}=-igf^{abc}\Gamma_{ij4}$.
 All functions $D_{ij}$ and $\Gamma_{ij4}$ are understood to be
 exact ones. The  main problem is to find the nonperturbative
expression for the  $\Gamma_{ij4}(r,p,q)$-function.

Since we do not know the exact expression for the three-gluon
vertex function, we try
a  nonperturbative
ansatz, which we  exploit in the following.
First the vertex function is
decomposed as
$$
\Gamma_{4ij}^{abc}(q,r,p)=\Gamma_{4ij}^{abc}(q,r,p)^{L}+
\Gamma_{4ij}^{abc}(q,r,p)^{T}\, ,\
\eqno{(5)}
$$
and its transverse and longitudinal parts are constructed independently.
In the following the longitudinal one is used in the form, which
  has been found in
Ref.[8]:
\setcounter{equation}{5}
\begin{eqnarray}
&&\Gamma_{4ij}^{abc}(q,\,r,\,p)^{L}=-igf^{abc}\left\{\delta_{ij}
(r_4-p_4)-\frac{1}{r^2-p^2}\left[\left(\frac{G(r)}{\r^2}-
\frac{F(r)}{r^2}\frac{r_4^2}{\r^2}\right)\right.\right.\nonumber\\
&&\hspace{2em}
-\left.\left.\left(\frac{G(p)}{\p^2}-\frac{F(p)}{p^2}\frac{p_4^2}
{\p^2}\right)\right][(\p\r)\delta_{ij}-p_ir_j](r_4-p_4)
\right.\nonumber\\
&&\hspace{2em}
+\left.\delta_{ij}\left(r_4\frac{F(r)}{r^2}-\frac{F(p)}{p^2}p_4\right)
+\frac{1}{q^2-r^2}\left(\frac{F(q)}{q^2}-\frac{F(r)}{r^2}\right)
q_ir_4(q-r)_j\right.\nonumber\\
&&\hspace{2em}
\left.+\frac{1}{p^2-q^2}\left(\frac{F(p)}{p^2}
-\frac{F(q)}{q^2}\right)(p-q)_ip_4q_j\right.\nonumber\\
&&\hspace{2em}
\left.-\frac{1}{r^2-p^2}\left(\frac{F(r)}{r^2}-\frac{F(p)}{p^2}\right)
r_4p_4(r-p)_4\delta_{ij}\right\}\,,\
\end{eqnarray}
and which is assumed to be
valid for any set of momenta including the soft ones.
Eq.(6) satisfies  the exact Slavnov-Taylor identity by using
the standard inverse gluon propagator
$$
{\cal {D}}_{ij}^{-1} (p) =\left(\delta_{ij}-\frac{p_{i}p_{j}}{\p^2}
\right)(p^2+G(p))+\left(1+\frac{F(p)}{p^2}\right)p_{4}^2
\frac{p_{i}p_{j}}{\p^2}\, ,\
\eqno{(7)}
$$
and more details of its properties can be found in Refs.[9,10].

\noindent
 The transverse part of the $\Gamma_{4ij}^{abc}(q,r,p)$-
function
  is not constrained by
the Slavnov-Taylor identity of Eq.(3),
 but it is introduced to cancel the leading infrared
singularities of the longitudinal vertex function (6).
It is given by
\setcounter{equation}{7}
\begin{eqnarray}
  \Gamma_{4ij}^{abc}(q,r,p)^{T} &=&
-igf^{abc}\left\{ \frac{- (\p\q) \dij + p_i  q_j}
{q^2-r^2} \left(\frac{F(q)}{q^2}-\frac{F(r)}{r^2}\right) r_4
\right.\nonumber\\
&&\hspace{2em}
\left.
-\frac{- (\r\q) \dij + r_j  q_i}{q^2-p^2}
\left(\frac{F(q)}{q^2}-
\frac{F(p)}{p^2}\right) p_4
\right\} \, ,\
\end{eqnarray}
and together with
 the transverse part of the function
$\Gamma_{44j}^{abc}(q,r,p)$
\setcounter{equation}{8}
\begin{eqnarray}
\Gamma_{44j}^{abc}(q,r,p)^{T}=
-igf^{abc}\left\{\frac{- (\p\r)q_j +(\p\q)r_j}
{q^2-r^2}\left(\frac{F(q)}{q^2}-\frac{F(r)}{r^2}\right)\right\}\, ,\
\end{eqnarray}
Eq.(8) is constrained by
 two identities
$$
\Gamma_{4{\mu}j}^{T}(q,r,p)r_{\mu}=0\,,\qquad
\Gamma_{4i{\mu}}^{T}(q,r,p)p_{\mu}=0\, .\
\eqno{(10)}
$$

In constructing the vertex function
$\Gamma_{4ij}$  it is necessary to impose that it is
well-behaved in the infrared limit, i.e. for
$q_4 = 0, \q \to 0$ (and similar for the momenta $p$ and $r$).
Indeed, it is the case with respect to $p$ and $r$ for the
longitudinal and transverse parts separately, but not for the infrared
limit in $q$.
However, the sum in Eq.(5) is infrared stable, as one can
easily verify. Therefore imposing this condition not only motivates
the inclusion of the transverse part Eq.(8),
 but it also fixes its magnitude
and (minimal) structure.

In order to determine
 $\Pi_{44}(0,|\q|)$  both the longitudinal (6) and transverse (8)
vertex functions are taken into account in
Eq.(4). In the following we restrict the calculation of
 $\Pi_{44}$  only up to the $O(g^3)$ term.
In this approximation
essential simplifications occur due to the different infrared behaviour
of the functions G(p) and F(p) in the static limit.
If required the function G(p) may be used as an infrared
cutoff (and e.g. be  identified with the magnetic mass), but
in the final stage of the calculation we set $G = 0$.
The dependence on the function F(p)
is evaluated in the infrared limit by keeping only the
$p_4=0$ mode in the Matsubara sum of Eq.(4) after identifying
the leading term of
the $\Pi_{44}$-function by
$\Pi_{44} = m^2=g^2 N T^2/3$.
The poles at $p_4=0$
 are regularized in the standard manner
when using  the temporal axial gauge
by the prescription:
$\sum_{p_4} 1/p_4^2 =0$.

\noindent
After the lenghty algebra is
performed we obtain
\setcounter{equation}{10}
\begin{eqnarray}
\Pi_{44}(0,|\q|)&=&\frac{g^2N}{3\beta^2}
+\frac{2g^2N}{\beta}\int\frac{d^3\p}{(2\pi)^3}\left\{
\frac{m^2 [\p^2 + (\p\q)]^2}{\p^2 (\p+\q)^2 [\p^2+m^2]
[(\p+\q)^2+m^2]}\right.\nonumber\\
&&\left.-\left(1-\frac{(\p\q)^2}{\q^2\p^2}\right)
\frac{2\q^2}{ \p^2 [(\p+\q)^2+m^2]}\right\}\, ,\
\end{eqnarray}
where all integrals have no divergencies and can be explicitly
evaluated.

The final result to order $g^3$
becomes
\setcounter{equation}{11}
\begin{eqnarray}
&&\Pi_{44}(0,|\q|)=m^2+\frac{g^2NmT}{2\pi}\left\{-\frac{1}{2}+
\frac{\pi}{8}\frac{\q^4}{m^3\sqrt{\q^2}}
\right.\\
&&\left.
+\left[ \frac{m^2-\q^2}
{m\sqrt{\q^2}}
-\frac{(\q^2+m^2)^2}{2m^3\sqrt{\q^2}}
\right] {\arctan}[\frac{\sqrt{\q^2}}{m}]+\frac{(\q^2+2m^2)^2}
{4m^3\sqrt{\q^2}}{\arctan}[\frac{\sqrt{\q^2}}{2m}]\right\}\,,\
\nonumber\end{eqnarray}
valid for real values of $\q$.
It is important to note that in the low momentum expansion
of $\Pi_{44}(0,|\q|)$  a positive
 cubic term ${|\q|}^3$ is present in Eq.(12)
- instead of a negative linear one in the case of the one-loop
approximation [11].  It originates from the product
of the longitudinal - longitudinal modes of ${\cal D}_{ij}$
in Eq.(4).

At $\q = 0$ the static self-energy is
\setcounter{equation}{12}
\begin{eqnarray}
\Pi_{44}(0)  & =& m^2+\frac{g^2NmT}{4\pi}
\\
&&
= \left[ \frac{g^2 N}{3} + \frac{3}{4\pi} (\frac{g^2N}{3})^{3/2} \right]
   T^2 \, ,\
\nonumber\end{eqnarray}
as obtained recently in Ref.[12].

For reference we quote here separately the $O(g^3)$
  contribution   due to the product of the
 longitudinal  and transverse modes of
 ${\cal D}_{ij}$,
$$
\Pi_{44}^{LT}(0,|\q|)=m^2+\frac{g^2NmT}{2\pi}
 \left[\frac{m^2-\q^2}
{m\sqrt{\q^2}}
{\arctan}[\frac{\sqrt{\q^2}}{m}]  -1 \right]            \, ,
\eqno{(14)}
$$
which coincides
 with the expression
(in the covariant Landau gauge)
 derived by Rebhan [1,2]
in the framework of the Braaten - Pisarski resummation scheme [3].

\bigskip
\leftline{\bf 3 The screening potential beyond the leading term}
\bigskip

 Applying linear response theory it is possible to relate
 $\Pi_{44}(0,|\q|)$
to the gluon-electric  potential.
For the colour-singlet case it is represented by the Fourier
transform [4]:
\setcounter{equation}{14}
\begin{eqnarray}
V(\r) &=&
  -\frac{N^2-1}{2N}  g^2(T)
\int\frac{d^3q}{(2\pi)^3}\frac{e^{i\q\r}}
{\q^2+\Pi_{44}(0,|\q|)}
\\
&&
 = \,  -\frac{N^2-1}{2N}
 \frac{g^2(T)}{2\pi^2  r}
\int \limits_{0}^{\infty}
\frac{  dq\, q\, \sin (qr)}{q^2+\Pi_{44}(0,q)}\, ,\
\nonumber\end{eqnarray}
where $r = |\r|$ and $q = |\q|$.

Considering
 large  distances  $r$ it is usually assumed
 that only the static $\Pi_{44}(0,|\q|=0)$-limit is essential
for evaluating Eq.(15), and that in this limit
the observable screening length is related to
 the electric mass $m_E^2=\Pi_{44}(0)$, as defined
 in the temporal axial
gauge.
Since the leading term for $g \to 0$
of this quantity (named as the
Debye mass) was found to be gauge-independent
it is the belief that the exponential  Debye screening is
also the result of  multi-loop calculations, although the momentum
dependence of the gluon self-energy is expected to have
 a rather complicated analytical
structure, e.g. branch points.
 Even for the case  that the only arising
singularities are poles, it cannot be excluded that they are
becoming complex ones
 as  solutions              of
$$
   \q^2+\Pi_{44}(0,\sqrt{\q^2})=0\,,\
\eqno{(16)}
$$
such that oscillations in the $r$ dependence of the potential are
resulting.

It was demonstrated in Refs.[6,7] that
 odd powers of $\sqrt{\q^2}$, with the branch point  at $\q^2=0$,
are responsible for
 the asymptotic power behaviour of $V(r)$ for large distance $r$
and   for high temperature.

In our case the situation
 is analogous to the one discussed in the paper [6],
 because of the term
 in $\Pi_{44}$ of Eq.(12), which is
 odd (cubic) in powers of $\sqrt{{\q}^2}$.
In order to facilitate the comparison with the one-loop result
the same notation as in Ref.[6] is used in the following.
The  convenient variables are $q=mz\,,\ x=mr$
and $t=g^2NT/ (8m) = 3 m /( 8 T)$, and a dimensionless
screening function $S(x,t)$ is defined by
$$
V(\r)=-\frac{N^2-1}{2N}\frac{g^2(T)}{4\pi r} S(x,t) \, ,\
\eqno{(17)}
$$
and
$$
 S(x,t) = \frac{2}{\pi} \int
\limits_{0}^{\infty}\frac{z  dz\, \sin(x z)}{[z^2 + 1+f(z)+ tz^3/2]}\,,\
\eqno{(18)}
$$
where the function $f(z)$ is obtained from Eq.(12),
$$
f(z)=  \frac{2 t}{\pi}[-1
+(1-4z^2-z^4)\frac{{\arctan}( z)}{z}+(z^2+2)^2
\frac{{\arctan}(z/2)}{2z}].
\eqno{(19)}
$$
Applying contour  integration (with the contour
 surrounding the first quadrant in the complex $z-$ plane)
 the leading term
in the high (but finite) temperature limit becomes
$$
 S(x,t) \simeq \frac{t}{\pi} \int
\limits_{0}^{\infty}\frac{ \,dy\, y^4 e^{- xy}}
{[(1 - y^2 + f(y))^2+ t^2y^6/4]}\,,\
\eqno{(20)}
$$
with $f(y)=f(-iz)$ on the first Riemann sheet.
Contributions arising from possible (complex) poles and
from cuts with imaginary   branch points
 may be neglected,
 since they vanish
exponentially.
The integral (20) is evaluated asymptotically
 by keeping only
the leading term for $y \simeq 0$
 in the denominator,
 since the other contributions become
exponentially small for large $x$, or are subleading for $ g \to 0$.
Thus
 a power-behaved  repulsive
potential
$$
V( r) {\vert}_{mr\rightarrow \infty}
 \simeq \frac{N^2-1}{8 \pi^2}\,\frac{3g^2(T)T}
{(rm)^6}  \,  \
\eqno{(21)}
$$
results.

The dependence of the exact $S(x,t)$ of  Eq.(18)
 on $x = m r$ for a fixed value of $t$, i.e. of
the coupling $g$ respectively,  is plotted in Fig.~1
for $t = 0.5$ (Fig. ~1a) and for $t = 1.0$ (Fig.~1b).
The dotted curve represents the exponential Debye form for
$t = 0$ at leading order, $S(x, 0) = \exp(-x)$.
The full curve is the result of our analysis with $f(z)$
of Eq.(19) and the dashed curve is from Eq.(14).

 One can see that after including the momentum
dependence in $\Pi_{44}$ as described above
 the exponential behaviour of $S(x,t)$
is significantly modified:
screening is increased for values of $x \le 3$ (that
means a positive correction to the Debye mass),
 but at larger
distances there is antiscreening due to the cubic momentum term
in $\Pi_{44}$
 with
$S(x,t) \simeq - 24 t/(\pi x^5)$. A similar behaviour has
 been discussed in
Ref.[6] in the case of the one-loop approximation
(for a comparison one may see Fig.~1 of Ref.[6]).

The dashed curve in Fig.~1 corresponds to the case given by Eq.(14),
which is analogous to the situation discussed in Refs.[1,2]
(for the gauge parameter fixed by $\alpha = 0$),
but without introducing a magnetic term.
No odd terms in the momentum $q$ are present.
The screening function $S(x,t)$
 tends to become
oscillatory around its zero value
\footnote{ But it has been pointed out  by A.~K.~Rebhan
that after introducing a small magnetic mass
 term in Eq.(14) as described
in [1,2] the corresponding function $S(x,t)$  does not
change its sign, i.e. it remains positive. }
; this behaviour is more transparent for increasing values of the
coupling as illustrated in Fig.~1b: for this purpose a rather extreme
value of $t = 1.0$ is chosen. For this large value even the
full curve shows two oscillations, but for $x > 7.0$
it remains negative and power-behaved. The dashed one
shows  oscillations for all $x > 10.0$, but which
are damped by $\exp( -x)$.

In order to test the reliability of our calculation with respect
to its infrared sensitivity we introduce a small magnetic mass term
in the transverse part of the gluon propagator by assuming
a non-vanishing, but constant value for the function $G$ of Eq.(2).
Numerically we find that the full curve in Fig.~1 remains
stable with respect to this modification. Indeed this should be the case,
since the cubic term in the momentum dependence of $\Pi_{44}$
 does not depend on the transverse part of
the gluon propagator, i.e. on the magnetic mass.

\bigskip

\bigskip
\leftline{\bf 4 Conclusion}
\bigskip

To summarize, we have analysed the static colour-singlet
potential $V(r)$ in QCD at high temperature. In the temporal
axial gauge and beyond leading order $V(r)$
  turns out
to be power-behaved $(\simeq 1/r^6)$  and repulsive (Eq.(21)).
Although the detailed shape of this potential is
gauge dependent,  the described qualitative features
may actually be independent of the chosen gauge, i.e.
as a result of the next-to-leading order treatment,
the positive cubic term in $q$ of $\Pi_{44}(0,\q)$  already
reflects the correct behaviour of
 the gluon self-energy for small momenta
$\q \simeq 0$.
We expect that this cubic term (contrary to the linear one
appearing in  the one-loop approximation)
remains infrared stable and survives under the further
resummations.
So the stated problem turns out to be more complicated
than for the usually considered ansatz with the Debye mass,
and more detailed investigations are desirable as a future task.

\begin{center}
{\bf Acknowledgements}
\end {center}

We are grateful to A.~K.~Rebhan for useful discussions.
R.~B. thanks J.~Kapusta for pointing out to him the relevance
of the temporal axial gauge.
O.~K.~K. would like to thank
the colleagues from the Department of Theoretical Physics of the
Bielefeld University for the kind hospitality, and acknowledges
partial support by
"Volkswagen - Stiftung".
This research is also supported in part by the EEC Programme
"Human Capital and Mobility", Network "Physics at High Energy
Colliders", contract CHRX-CT93-0537 (DG 12 COMA).

\newpage

\begin{center}
{\bf References}
\end{center}

\renewcommand{\labelenumi}{\arabic{enumi}.)}
\begin{enumerate}

\item{ A.~K.~Rebhan, Phys. Rev. {\bf D48} (1993) R3967.  }

\item{A.~K.~Rebhan,
  "Gauge-independent extraction of the next-to-leading
   -order Debye mass from the gluon propgator",
    contribution at the {\sl 3rd Workshop on Thermal Field Theories and
    their Applications}, Banff, Canada, August 1993,
     preprint BI-TP 93/52.}

\item{E.~Braaten and R.~D.~Pisarski,
 Phys. Rev. Lett. {\bf 64} (1990) 1338;
   Nucl. Phys. {\bf B337} (1990) 569.}

\item{K.~Kajantie and J.~Kapusta, Ann. Phys. (N.Y.)
 {\bf 160} (1985) 477.}

\item{ J.~I.~Kapusta, {\sl Finite Temperature Field Theory},
   Cambridge University Press, Cambridge, England, 1989.}

\item{C.~Gale and J.~Kapusta, Phys. Lett. {\bf B198} (1987) 89.  }

\item{J.~Kapusta and T.~Toimela. Phys. Rev. {\bf D37} (1988) 3731.}

\item{ O.~K.~Kalashnikov, JETP Lett. {\bf 39} (1984) 405.}

\item{ O.~K.~Kalashnikov,  in
   {\sl Quantum Field Theory and Quantum Statistics}
   (I.~A.~Batalin et al., eds.),  Adam Hilger, Bristol, 1987.}

\item{ O.~K.~Kalashnikov, Fortschr. Phys. {\bf 32} (1984) 525.}

\item{ O.~K.~Kalashnikov and V.~V.~Klimov,
   Sov. J. Nucl. Phys.    {\bf 33} (1981) 443;
 H.~A.~Weldon, Phys. Rev. {\bf D26} (1982) 1394.}

\item{ O.~K.~Kalashnikov, "The nonperturbative equation for the
   infrared $\Pi_{44}(0)$-limit in the temporal axial gauge",
   preprint BI-TP 93/77 (December 1993). }

\end{enumerate}
\bigskip

\bigskip
\begin{center}
{\bf Figure Caption}
\end{center}
\bigskip

Fig.~1 \quad
 The screening function $S(x,t)$ as a function of
$x$ for two values of  $t=0.5$ (Fig.~1a) and $t=1.0$
(Fig.~1b) as described in the text. The change in scales
(logarithmic and linear) has to be noted.

\end{document}